\documentclass[10pt]{IEEEtran}
\IEEEoverridecommandlockouts
\usepackage{cite}
\usepackage{amsmath,amssymb,amsfonts}
\usepackage{algorithmic}
\usepackage{adjustbox}
\usepackage{graphicx}
\usepackage{textcomp}
\usepackage{comment}
\usepackage{setspace}
\usepackage{stfloats}
\usepackage{xcolor}
\usepackage{multirow}
\usepackage{hyperref}
\usepackage{tcolorbox}
\tcbuselibrary{skins, breakable}
\usepackage{xcolor}
\usepackage{booktabs}
\usepackage{listings}
\usepackage{makecell}
\usepackage{colortbl}
\usepackage{array}
\usepackage{adjustbox}
\usepackage{inconsolata}
\usepackage[font=footnotesize,skip=0pt]{caption}

\tcbuselibrary{listingsutf8}

\newtcolorbox[auto counter, number within=section]{mybox}[2][]{
    colback=gray!10!white, colframe=gray!35,
    fonttitle=\bfseries, coltitle=black,
    title=#2, #1
}

\lstdefinestyle{python}{
    language=Python,
    basicstyle=\ttfamily\footnotesize,
    keywordstyle=\color{blue},
    commentstyle=\color{green!60!black},
    stringstyle=\color{red},
    numbers=left,
    numberstyle=\tiny\color{gray},
    stepnumber=1,
    numbersep=5pt,
    backgroundcolor=\color{white},
    frame=single,
    rulecolor=\color{black!30},
    tabsize=4,
    captionpos=b,
    breaklines=true,
    breakatwhitespace=true,
    showstringspaces=false
}

\def\BibTeX{{\rm B\kern-.05em{\sc i\kern-.025em b}\kern-.08em
    T\kern-.1667em\lower.7ex\hbox{E}\kern-.125emX}}
\lstdefinestyle{verilog}{
    basicstyle=\ttfamily\scriptsize,  % Reduced font size
    keywordstyle=\color{blue},
    commentstyle=\color{green!60!black},
    stringstyle=\color{red},
    numbers=left,
    numberstyle=\tiny\color{gray},
    stepnumber=1,
    numbersep=5pt,
    backgroundcolor=\color{white},
    frame=single,
    rulecolor=\color{black!30},
    tabsize=4,
    captionpos=b,
    breaklines=true,
    breakatwhitespace=true,
    showstringspaces=false
}

\usepackage[compact]{titlesec}

\titlespacing\section{0pt}{0.3\baselineskip}{0.2\baselineskip}
\titlespacing\subsection{0pt}{0.2\baselineskip}{0.1\baselineskip}
\titlespacing\subsubsection{0pt}{0.15\baselineskip}{0.1\baselineskip}

\usepackage{enumitem}

\setlist[itemize]{leftmargin=*}%
\setlist[enumerate]{leftmargin=*}%
    \usepackage{fancyhdr}
\fancypagestyle{firstpage}{
   \fancyhf{} % clear all header and footer fields
   \fancyhead[C]{To appear at 2026 IEEE International Joint Conference on Neural Networks (IJCNN) Proceedings, Maastricht, Netherlands} % Centered header
}
\begin{document}

% \title{PennyLang: Pioneering LLM-Based Quantum Code Generation with a Novel PennyLane-Centric Dataset}

% \title{A PennyLane-centric Dataset to enhance LLM-based Quantum Code Generation using Retrieval-Augmented Generation}

\title{A PennyLane-Centric Dataset to Enhance LLM-based Quantum Code Generation using RAG
\vspace{-10pt}
}

% \author{\IEEEauthorblockN{Anonymous Authors}}

\author{
\IEEEauthorblockN{Abdul Basit\textsuperscript{1}, Nouhaila Innan\textsuperscript{1,2}, Muhammad Haider Asif\textsuperscript{1,2}, Minghao Shao\textsuperscript{1}, \\Muhammad Kashif\textsuperscript{1,2}, Alberto Marchisio\textsuperscript{1,2}, Muhammad Shafique\textsuperscript{1,2}\\}
\IEEEauthorblockA{\textit{\textsuperscript{1}eBRAIN Lab, Division of Engineering} \textit{New York University (NYU) Abu Dhabi}, Abu Dhabi, UAE}\\
\textit{\textsuperscript{2}Center for Quantum and Topological Systems (CQTS), NYUAD Research Institute, New York University Abu Dhabi}, UAE\\
\{abdul.basit, nouhaila.innan, ma8183, shao.minghao, muhammadkashif, alberto.marchisio, muhammad.shafique\}@nyu.edu 
\vspace{-10pt}
}

\maketitle
\thispagestyle{empty} %page number
\thispagestyle{firstpage}

\begin{abstract}
Large Language Models (LLMs) offer powerful capabilities in code generation, natural language understanding, and domain-specific reasoning. Their application to quantum software development remains limited, in part because of the lack of high-quality datasets both for LLM training and as dependable knowledge sources. To bridge this gap, we introduce \textit{PennyLang}, an off-the-shelf, high-quality dataset of 3,347 PennyLane-specific quantum code samples with contextual descriptions, curated from textbooks, official documentation, and open-source repositories. Our contributions are threefold: (1) the creation and open-source release of \textit{PennyLang}, a purpose-built dataset for quantum programming with PennyLane; (2) a framework for automated quantum code dataset construction that systematizes curation, annotation, and formatting to maximize downstream LLM usability; and (3) a baseline evaluation of the dataset across multiple open-source and commercial models, including ablation studies, all conducted within a retrieval-augmented generation (RAG) pipeline. Using PennyLang with RAG substantially improves performance: for example, Qwen 7B’s success rate rises from 8.7\% without retrieval to 41.7\% with full-context augmentation, and LLaMa 4 improves from 78.8\% to 84.8\%, while also reducing hallucinations and enhancing quantum code correctness. Moving beyond Qiskit-focused studies, we bring LLM-based tools and reproducible methods to PennyLane for advancing AI-assisted quantum development.
\end{abstract}
\begin{IEEEkeywords}
LLM, Quantum Computing, RAG
\end{IEEEkeywords}
\section{Introduction}
Quantum computing has emerged as a transformative technology with the potential to solve complex problems that are infeasible for classical systems~\cite{feynman1982simulating, preskill2018nisq, kashif:2022}. Breakthroughs such as Google's demonstration of ``quantum supremacy''~\cite{arute2019quantum}, where a quantum processor performed a calculation in seconds that would take thousands of years on a classical supercomputer, highlight its disruptive capabilities in domains ranging from cryptography and quantum chemistry to machine learning and optimization~\cite{montanaro2016quantum, zaman2023survey, innan2024qfnn, el2024quantum, innan2024quantum1, zaman2024po, innan2024financial, kashif2023impact,pathak2024resource,el2024advqunn,10651123, kashif:2025computational, marchisio2025q, kashif2025evaluating, innan2025quav}.

\textbf{Research Challenges}: Despite these advances, programming quantum systems is challenging due to the specialized nature of existing frameworks and benchmarks, hindering widespread adoption~\cite{moll2018quantum, Wang_2024}. PennyLane is a leading open-source Python framework designed for hybrid quantum-classical computing, allowing seamless integration of quantum circuits with machine learning workflows~\cite{bergholm2018pennylane}. However, unlike the Qiskit framework~\cite{aleksandrowicz2019qiskit, kashif:qiskit}, which benefits from dedicated AI-driven code assistants~\cite{qiskitAssistant}, PennyLane lacks equivalent tools to assist developers in writing optimized quantum code.

In parallel, the field of Natural Language Processing (NLP) has seen rapid advancements in Large Language Models (LLMs), which now excel at reasoning, contextual understanding, and automated code generation~\cite{brown2020language, chen2021evaluating, li2022competition}. These models have shown success in classical programming~\cite{codex2021, codeLlama}, improving developer productivity and reducing syntax and logic errors. However, their application in quantum programming, especially within the PennyLane ecosystem, remains underexplored. This motivates our work: curating a high-quality, PennyLane-centric dataset and establishing a structured methodology for refinement, annotation, and evaluation, ultimately enabling LLM-based quantum code generation and debugging.

\textbf{State-of-the-Art and Limitations:}  
LLMs have been extensively benchmarked for classical programming~\cite{nijkamp2023codegen, fried2022incoder}, but their performance in quantum domains remains less explored. Early efforts such as Qiskit HumanEval and Qiskit Code Assistant~\cite{qiskitAssistant} adapt LLMs to quantum programming but focus on Qiskit’s hardware-centric paradigm. PennyLane, by contrast, targets variational quantum algorithms and quantum machine learning~\cite{schuld2018supervised}, requiring dedicated LLM support and a specialized training corpus. A major barrier is the scarcity of suitable data: quantum code is scattered across papers, forums, repositories, and documentation~\cite{Ren2024}, and often lacks the contextual annotations necessary for LLMs to link quantum operations with their intended use.

\begin{figure}[t]
    \centering
    \includegraphics[width=1.02\linewidth]{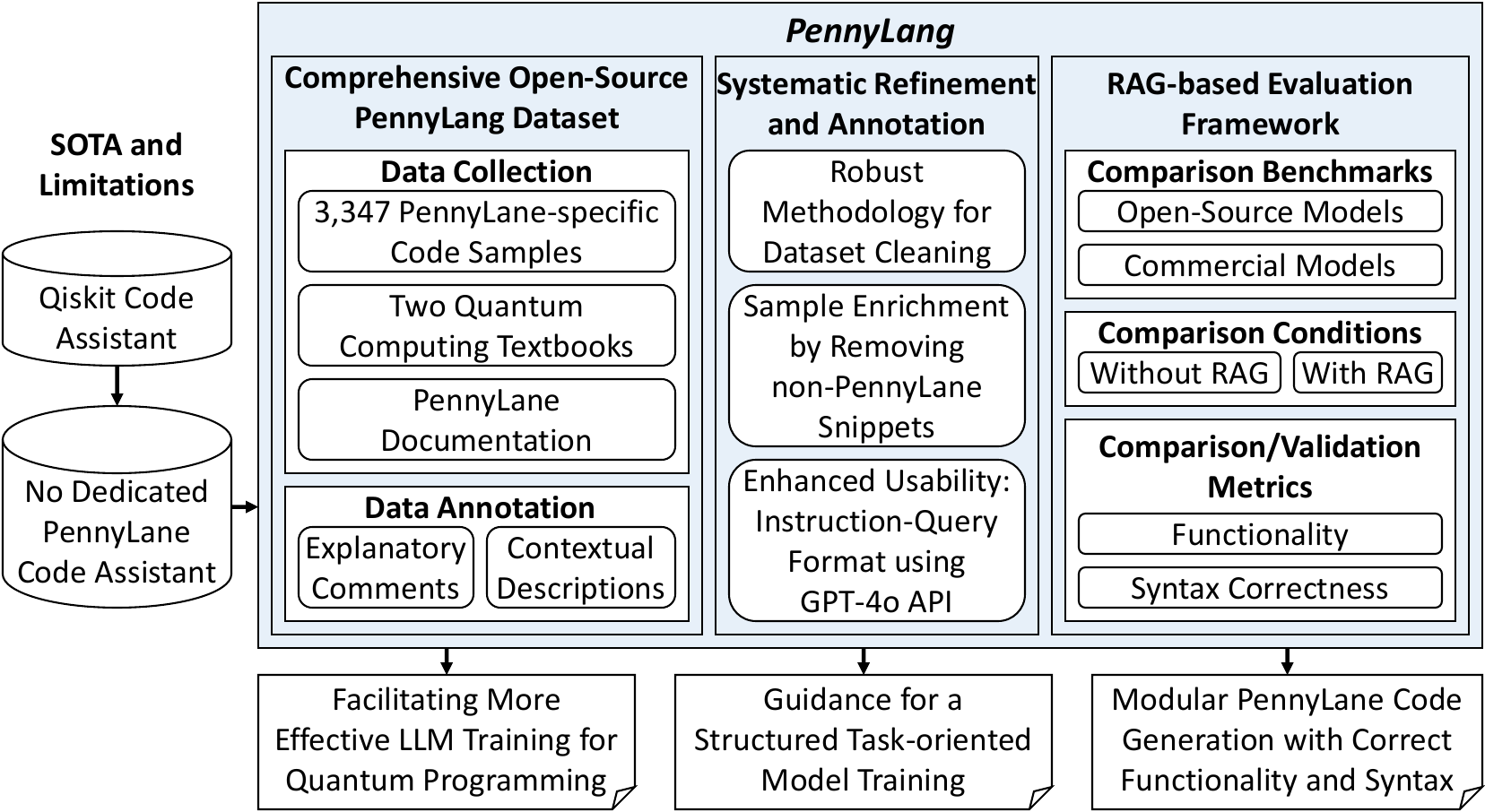}
    \caption{Overview of novel contributions in this work.
}
    \label{fig:novelcontribution}
    \vspace{-2pt}
\end{figure}

\textbf{Our Approach and Contributions:}  
In this work, we introduce \textit{PennyLang}, a large-scale, high-quality dataset specifically designed for PennyLane code generation, accompanied by a rigorous evaluation pipeline. We implement a baseline Retrieval-Augmented Generation (RAG) framework to assess the dataset’s effectiveness across multiple models. Our key contributions, shown in Fig.~\ref{fig:novelcontribution}, consist of:

\begin{enumerate}
    \item \textit{Comprehensive PennyLane Dataset:}  
    Curated 3,347 PennyLane-specific code samples from GitHub repositories, quantum computing textbooks, and official PennyLane documentation, each enriched with contextual descriptions and annotations to support instruction tuning.

    \item \textit{Systematic Data Refinement and Annotation:} Developed a robust pipeline for dataset cleaning, metadata enrichment, and conversion of code snippets into instruction–query pairs using the GPT-4o API, ensuring compatibility with LLM training workflows.

   \item \textit{RAG-based Evaluation Framework:} Designed a RAG evaluation framework to assess the dataset’s impact on model performance. Framework was evaluated on seven LLMs, including both open-source and commercial models under both retrieval-augmented and non-augmented settings, scoring outputs on functional correctness.

\end{enumerate}

\textbf{Summary of Key Results:}
Retrieval augmentation with PennyLang yields substantial gains for smaller open-source models: Qwen 7B’s success rate rises from 8.71\% to 41.66\%, %($\approx$32.95\% absolute increase), 
and LLaMa 4 improves from 78.78\% to 84.84\%. %($\approx$7.7\% relative increase). 
In contrast, stronger baselines (commercial models) do not benefit significantly from naive full-context retrieval. For example, the best performance is achieved by Claude 3.5 Sonnet without retrieval (95.07\%), with full retrieval resulting in slight degradation. Detailed analysis is presented in Section~\ref{sec:Evaluation}.

\section{Background and Related Work \label{sec2}}

\begin{table*}[ht!]
    \centering
    \caption{Summary of Qiskit and PennyLane Datasets for LLM Training}
    \begin{adjustbox}{max width=\linewidth}
    \begin{tabular}{|>{\columncolor{gray!10}}l|l|l|c|l|l|}
        \hline
        \rowcolor{gray!25}
        \textbf{Dataset Name} & \textbf{Framework} & \textbf{Source} & \textbf{Samples Available} & \textbf{Dataset Type} & \textbf{Format} \\
        \hline\hline
        
        % Qiskit Datasets Section
        \multicolumn{6}{|c|}{\cellcolor{blue!15}\textbf{\textsc{Qiskit Framework Datasets}}} \\
        \hline
        
        {Qiskit Textbook Dataset}~\cite{qiskit_textbook} & 
        \texttt{Qiskit} & 
        Qiskit.org Textbook & 
        \textbf{Multiple} & 
        Structured Qiskit Code & 
        Python Scripts \\
        \hline
        
        {IBM Quantum Challenge Data}~\cite{ibm_quantum_challenge} & 
        \texttt{Qiskit} & 
        IBM Quantum Experience & 
        \textbf{Derived} & 
        Instruction-Response Format & 
        JSON/Notebook \\
        \hline
        
        {Qiskit Code Repository}~\cite{qiskit_github} & 
        \texttt{Qiskit} & 
        GitHub (Qiskit AI repos) & 
        \textbf{Thousands} & 
        Raw Qiskit Code Samples & 
        Python Scripts \\
        \hline
        
        {Q-Gen Quantum Circuit}~\cite{qiskit_kaggle} & 
        \texttt{Qiskit} & 
        Kaggle (Possible) & 
       \textbf{309} & 
        Pretrained Benchmark Data & 
        CSV/JSON \\
        \hline
        
        {QDataSet}~\cite{qiskit_instruction_data} & 
        \texttt{Qiskit} & 
        GitHub & 
        \textbf{52} & 
        Quantum ML Data & 
        Various \\
        \hline
        
        {QCircuitNet}~\cite{mitQuantumData} & 
        \texttt{Qiskit} & 
        Research Paper & 
        \textbf{4,800} & 
        Quantum Algorithm Design Benchmark & 
        Python/QASM \\
        \hline\hline
        
        % PennyLane Datasets Section
        \multicolumn{6}{|c|}{\cellcolor{orange!15}\textbf{\textsc{PennyLane Framework Datasets}}} \\
        \hline
        
        {PennyLane Documentation}~\cite{pennylaneDocs} & 
        \texttt{PennyLane} & 
        PennyLane.ai Docs & 
        \textbf{Various} & 
        Raw PennyLane Code Samples & 
        Python Scripts \\
        \hline
        
        {PennyLane GitHub Repos}~\cite{pennylaneaiGitHub} & 
        \texttt{PennyLane} & 
        GitHub Open-Source & 
        \textbf{Scattered} & 
        Unstructured Quantum Code & 
        Python Scripts \\
        \hline
        
        \rowcolor{green!20}
        {\textbf{PennyLang (Ours)}} & 
        \texttt{\textbf{PennyLane}} & 
        \textbf{Open-Source Dataset} & 
        \textcolor{blue}{\textbf{3,347}} & 
        \textbf{Curated and Structured Dataset} & 
        \textbf{JSON file} \\
        \hline
        
    \end{tabular}
    \end{adjustbox}
    \label{tab:qiskit_pennylane_datasets}
\end{table*}

The intersection of LLMs and quantum computing is an emerging research domain. While LLMs have demonstrated impressive capabilities in NLP~\cite{Shao_2024} and \textit{automated code generation}~\cite{jiang2024surveylargelanguagemodels}, their application in quantum software development remains underexplored. Most existing studies focus on classical programming languages or quantum frameworks such as \textit{Qiskit}~\cite{qiskitAssistant}, with little attention to the PennyLane framework~\cite{bergholm2018pennylane, basit2025pennycoder, basit2025qhackbench}, despite its popularity in hybrid quantum-classical workflows.
This section presents an in-depth review of related work in three key areas:

\begin{enumerate}
    \item LLMs for Code Generation in Classical Programming
    \item LLMs for Quantum Programming
    \item Datasets for Training LLMs in Quantum Code Generation (summarized in Table~\ref{tab:qiskit_pennylane_datasets})
\end{enumerate}

We then summarize existing contributions and highlight our novel approach in developing a \textit{PennyLane-centric dataset} to advance AI-driven quantum programming, as shown in Table~\ref{tab:comparison}.

\subsection{LLMs for Code Generation in Classical Programming}
Transformer-based language models have revolutionized automated code generation, significantly advancing classical software development. OpenAI’s Codex~\cite{codex2021}, trained on large-scale public code repositories, enabled natural language-to-code translation, setting a new standard for programming assistants. CodeBERT~\cite{feng2020codebert} further improved source code understanding through bidirectional transformers, while InCoder~\cite{fried2022incoder} introduced multi-turn interactive capabilities for code infilling and synthesis. AlphaCode~\cite{li2022competition} demonstrated that LLMs could generate solutions for competitive programming problems, reaching near-human performance. To rigorously evaluate these models, benchmarks such as HumanEval~\cite{chen2021evaluating}, MBPP~\cite{austin2021mbpp}, and CodeXGLUE~\cite{lu2021codexglue} have been proposed, each offering complementary perspectives on correctness, diversity, and execution fidelity. These benchmarks collectively emphasize the importance of well-curated, structured datasets in improving LLM-driven code generation.
\subsection{LLMs for Quantum Programming}
In comparison, quantum programming has seen limited adoption of LLMs, with most efforts focused on the Qiskit framework. IBM’s Qiskit Code Assistant~\cite{qiskitAssistant} supports Qiskit programs' debugging and optimizations, while \cite{li2023adaptingpretrainedlanguagemodels} adapted BERT-based models for enhanced quantum code comprehension. CodeLlama-Quantum \cite{codeLlama} extended Meta’s CodeLlama to address Qiskit-specific use cases. Despite these advances, key limitations persist. The field lacks large, high-quality datasets for training and evaluation, and quantum circuits demand deeper reasoning due to entanglement, measurement, and complex control logic. Furthermore, hybrid programming paradigms like PennyLane interleave classical and quantum instructions, requiring models to understand multi-modal code patterns and contextual dependencies. To address these challenges, we propose \textit{PennyLang}, the first PennyLane-centric dataset, supported by an evaluation pipeline that integrates RAG to improve code relevance and contextual grounding. Our contributions advance {AI-assisted quantum programming} by enabling LLMs to provide {high-quality, domain-specific assistance} in PennyLane.

\begin{figure*}[t!]
    \centering
    \includegraphics[width=\linewidth]{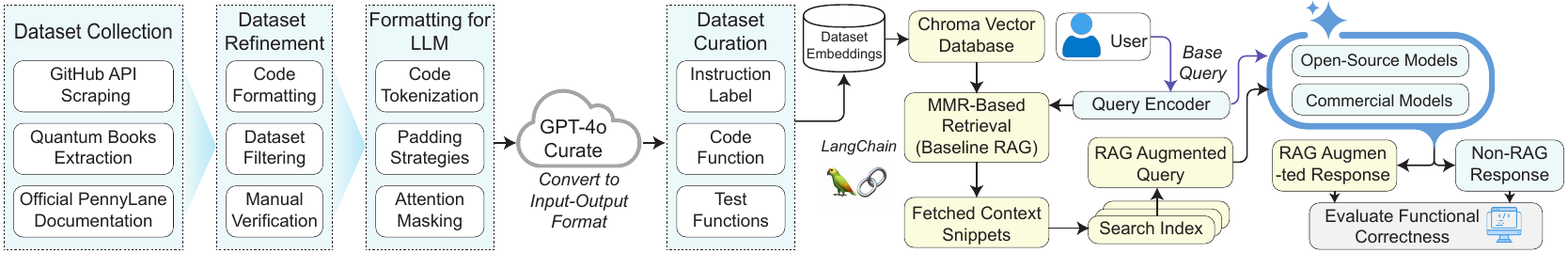}
    \caption{ \small \textbf{Methodology for PennyLang Quantum Code Generation and Evaluation.}
    We collect and refine quantum code from GitHub, textbooks, and PennyLane documentation, then use GPT-4o to convert cleaned snippets into instruction–query pairs (with code and tests). These examples are embedded in Chroma, and LangChain performs MMR-based retrieval to produce RAG-augmented versus vanilla prompts. Both open-source and commercial models are evaluated on functional correctness to quantify the benefit of the PennyLang dataset.}
    \label{fig:currentmethodology}
\end{figure*}
\label{sec:methodology}
\color{black}

\begin{table}[t]
    \centering
    \caption{Comparison: Related Work \& Our Contributions}
    \begin{adjustbox}{max width=\linewidth}
    \begin{tabular}{|c|c|c|}
        \hline
        \rowcolor{gray!25}
        \textbf{Research Area} & \textbf{Existing Works} & \textbf{Our Contributions} \\
        \hline\hline
        
        \makecell[c]{\textbf{LLMs for}\\ \textbf{Code Generation}} & 
        \makecell[c]{Codex~\cite{codex2021}\\ CodeBERT~\cite{feng2020codebert}} & 
        \makecell[c]{\textbf{PennyLane-specific}\\ \textbf{dataset}} \\
        \hline
        
        \makecell[c]{\textbf{LLMs for Quantum}\\ \textbf{Programming}} & 
        \makecell[c]{Qiskit Assistant~\cite{qiskitAssistant}\\ \cite{li2023adaptingpretrainedlanguagemodels}} & 
        \makecell[c]{\textbf{Hybrid quantum-}\\ \textbf{classical dataset}} \\
        \hline
        
        \makecell[c]{\textbf{Quantum Code}\\ \textbf{Datasets}} & 
        \makecell[c]{QCircuitNet~\cite{mitQuantumData}\\ QDataSet~\cite{qiskit_instruction_data}} & 
        \makecell[c]{\textbf{First PennyLane}\\ \textbf{dataset}} \\
        \hline
        
    \end{tabular}
    \end{adjustbox}
    \label{tab:comparison}
\end{table}

\section{PennyLang Framework \label{sec3}}
\subsection{Methodology for Data Collection and Refinement}

Fig.~\ref{fig:currentmethodology} illustrates the methodology employed to construct a {high-quality dataset of PennyLane code samples}, ensuring that each snippet is {genuinely relevant, well-structured, and accurately annotated}. Our data collection strategy focused on extracting, filtering, and refining code snippets from multiple sources, including {GitHub repositories, quantum computing books, and official PennyLane documentation}.

\subsubsection{Data Collection from GitHub Repositories}
To systematically gather PennyLane-specific code, we utilized the {GitHub API} to search for publicly accessible repositories with permissive {open-source licenses} (e.g., MIT, Apache 2.0, BSD). The selection criteria required that the term \texttt{"PennyLane"} be present in either the {repository name, description, or README file}. To {maintain dataset quality and avoid redundancy}, we followed these constraints:  

\begin{itemize}
    \item Only repositories from the \textbf{main branch at the latest commit} were considered.
    \item \textbf{Forked repositories} were systematically excluded.
    \item Only Python (\texttt{.py}) files were extracted for further analysis.
    \item Large, \textbf{autogenerated files} (e.g., test logs, package files) were excluded.
\end{itemize}

Since PennyLane maintains {backward compatibility} and rarely deprecates features~\cite{bergholm2018pennylane}, we did not impose a {time constraint} on data collection, ensuring a {broad dataset scope}. Additionally, introductory comment blocks containing {license information, author details, or unrelated metadata} were removed to retain only {executable and relevant code}. This curation process resulted in a {high-fidelity dataset} of PennyLane-specific implementations. In total, {$1{,}321$ samples} were collected from the official {PennyLaneAI GitHub repository}~\cite{pennylaneaiGitHub}.

\subsubsection{Extracting Quantum Code from Books}
In addition to GitHub, we extracted {PennyLane-related code snippets} from two quantum computing books, Quantum Machine Learning: An Applied Approach ~\cite{ganguly2021quantum} and A Practical Guide to Quantum Machine Learning and Quantum Optimization~\cite{combarro2023practical}. 
These books provide {theoretical foundations} and {applied examples} of PennyLane-based quantum machine learning (QML) and variational quantum algorithms (VQAs).  
The {extraction process} involved:

\begin{enumerate}
    \item Identifying \textbf{all PennyLane-related code snippets}.
        \item \textbf{Diversity Assurance}: The dataset includes a {broad spectrum of PennyLane use cases}, from {basic gate operations} to {advanced quantum optimization problems}.
    \item Extracting \textbf{only the relevant code} while discarding unrelated sections.
    \item \textbf{Manual verification} with explanations from the books, converting descriptive text into \textbf{structured comments}.
\end{enumerate}

Unlike structured repositories, {book content lacked a consistent format} for where explanations were placed (before or after code). Thus, {manual review} was necessary to ensure that {annotations correctly corresponded} to each snippet.

\subsubsection{Scraping Code from PennyLane Documentation}
The official PennyLane documentation~\cite{pennylaneDocs} is a comprehensive resource covering topics such as quantum circuits, variational algorithms, operators and templates, hybrid quantum–classical optimization, and quantum chemistry applications. To enrich our dataset, we systematically extracted all code snippets from the official website, including those embedded in tutorials and API references. Documentation text accompanying each example was manually converted into inline comments, ensuring every sample is self-contained and context-rich. This approach improves LLM interpretability, enabling models trained on the dataset to produce both functional code and explanatory text.

\subsubsection{Dataset Summary and Statistics}
The final dataset containing {$3{,}347$ PennyLane code samples} is shown in Table~\ref{tab:dataset}.

\begin{table}[h]
    \centering
    \caption{Dataset Composition}
    \begin{adjustbox}{max width=.9\linewidth}
    \begin{tabular}{|l|c|}
        \hline
        \textbf{Source} & \textbf{Samples Collected} \\
        \hline\hline
        
        {Unofficial GitHub repositories} & 1,952 \\
        {GitHub (PennyLaneAI repository)} & 1,321 \\
        {Official PennyLane Documentation} & 53 \\
        {Quantum Computing Books} & 21 \\       
        \hline
        \textbf{Total Dataset Size} & \textbf{3,347} \\
        \hline
    \end{tabular}
    \end{adjustbox}
    \label{tab:dataset}
\end{table}

Each sample varies in {token length}, depending on the {tokenizer used}. This dataset is now {suitable for enhancing LLMs for PennyLane-based quantum programming}.

\subsubsection{Ensuring Data Integrity and Diversity}
To keep {dataset integrity}, we applied multiple {quality control measures}:

\begin{enumerate}
    \item \textbf{Duplicate Removal}: Identical code samples from different sources were eliminated.
    \item \textbf{Code Formatting}: Samples were formatted consistently to follow \textbf{PEP 8 style guidelines}~\cite{van2001pep8}.
    \item \textbf{Manual Review}: A {human-in-the-loop verification} was conducted to assess annotation accuracy.
\end{enumerate}

These efforts ensure that the dataset is {rich, well-structured, and optimized for AI model training}.

\subsubsection{Implications for AI-Driven PennyLane Code Generation}
The collected dataset {significantly enhances} AI-driven PennyLane code assistance in multiple ways:

\begin{itemize}

    \item \textbf{Improving Model Performance}: High-quality annotations ensure models {understand both syntax and context}.
    \item \textbf{Facilitating Explainable AI}: Models trained on PennyLang can {generate explanatory comments} alongside code.
\end{itemize}

This dataset serves as a \textit{foundation for future research} in \textit{AI-assisted quantum programming for PennyLane}.

\subsection{Tokenization and Formatting for LLM}
This section details the preprocessing pipeline for preparing our dataset for (LLM), focusing on {tokenization, padding strategies, and attention masking} to optimize input handling within the transformer architecture.

\subsubsection{Padding Strategies and Tokenizer Variations}
Padding strategies differ across tokenizers:

\begin{itemize}
    \item \textbf{Token-Specific Padding}: Some tokenizers have a {dedicated padding token}, automatically appended to shorter sequences. For example, Hugging Face’s Tokenizer uses \texttt{"<pad>"} by default~\cite{huggingfaceTransformers}.
    \item \textbf{Manually Defined Padding}: If a tokenizer lacks an explicit padding token, a placeholder token (e.g., \texttt{"0"} or \texttt{"<mask>"}) must be manually assigned.
    \item \textbf{Left vs. Right Padding}: Left padding is commonly used in {autoregressive models}~\cite{transformerVaswani}, ensuring that padding tokens do not interfere with causal attention.
\end{itemize}

In our pipeline, we employ left padding as it aligns with the causal decoder-only transformer architectures used in models like GPT-4o Mini, Claude 3.5 Sonnet, Gemini 2.5 Flash, Claude Haiku, and GPT-5 Mini. This ensures that masked self-attention restricts predictions to depend only on preceding tokens, maintaining auto-regressive consistency.

\subsubsection{Applying Attention Masking}
To improve {transformer efficiency}, an {attention mask}~\cite{attentionMechanism} was applied post-padding. This mask helps the model distinguish between {valid input tokens} and {padding tokens}, ensuring that computation is focused on meaningful data. Table~\ref{tab:tokenization} shows the {attention mask matrix} applied to a batch of tokenized samples. Here, the \texttt{"<pad>"} tokens are assigned 0s, instructing the model to ignore them during processing.

\begin{table}[htpb]
    \centering
    % \scriptsize
    \caption{Tokenized Input and Attention Masking.}
    \renewcommand{\arraystretch}{0.8}
    \begin{adjustbox}{max width=\linewidth}
    \setlength{\tabcolsep}{3pt}
    \begin{tabular}{|c|c|c|c|c|c|}
        \hline
        \textbf{Token} & \textbf{Mask} & \textbf{Token} & \textbf{Mask} & \textbf{Token} & \textbf{Mask} \\
        \hline
        \texttt{"<pad>"} & 0 & \texttt{"def"} & 1 & \texttt{"circuit"} & 1 \\
        \texttt{"<pad>"} & 0 & \texttt{":"} & 1 & \texttt{"("} & 1 \\
        \texttt{")"} & 1 & \texttt{"return"} & 1 & \texttt{"qml.expval"} & 1 \\
        \hline
    \end{tabular}
    \label{tab:tokenization}
    \end{adjustbox}
\end{table}

\begin{itemize}
    \item \textbf{Attention Mask Structure}:  
    A binary mask is assigned to each token:
    \begin{itemize}
        \item \texttt{1} → Meaningful tokens to be processed by the model.
        \item \texttt{0} → Padding tokens ignored by the transformer.
    \end{itemize}
    \item \textbf{Impact}:  
    Self-attention~\cite{transformerVaswani} ensures that padding tokens {do not influence attention computations}, thereby {reducing unnecessary computations} and preventing misleading dependencies.
\end{itemize}

% _______________________________________________________________
\subsection{Methodology for Dataset Evaluation}

\subsubsection{Retrieval-Augmented Generation (RAG) Framework}

To improve PennyLane code generation by reducing hallucinations and outdated syntax, and to better align outputs with PennyLane best practices, we adopt a retrieval-augmented generation (RAG) \cite{RAG} approach. We implement the pipeline in LangChain with modular components for document ingestion, vector embedding, retrieval, and prompt construction. The PennyLang dataset, formatted as instruction–response pairs, is embedded using \texttt{OpenAIEmbeddings} and stored in a \texttt{Chroma} vector database for efficient similarity search. At inference, user queries are embedded, relevant context is retrieved, and the combined prompt is sent to the target LLM. Table~\ref{tab:langchain_components} summarizes the key PennyLang RAG framework components, including their role in dataset loading, embedding storage, prompt formatting, retrieval chaining, and interfacing with multiple LLM backends.

\begin{table}[ht!]
    \centering
    %\scriptsize
    \caption{PennyLang Components Used in the RAG-based Evaluation Framework}
    \begin{adjustbox}{max width=\linewidth}
    \begin{tabular}{|l|}
        \hline
        \textbf{PennyLang RAG Framework Components} \\
        \hline\hline
        
        \texttt{\textbf{DirectoryLoader}} \textit{(Document Loader)} \\
        \quad $\triangleright$ Loads PennyLane dataset samples as text files from directory \\
        \hline
        
        \texttt{\textbf{Chroma}} \textit{(Vector Database)} \\
        \quad $\triangleright$ Stores embeddings for retrieval using similarity search \\
        \hline
        
        \texttt{\textbf{OpenAIEmbeddings}} \textit{(Embedding Model)} \\
        \quad $\triangleright$ Converts text into vector embeddings for indexing and retrieval \\
        \hline
        
        \texttt{\textbf{ChatPromptTemplate}} \textit{(Prompt Formatting)} \\
        \quad $\triangleright$ Defines structured prompts for AI models to ensure consistency \\
        \hline
        
        \texttt{\textbf{create\_retrieval\_chain}} \textit{(Retrieval Pipeline)} \\
        \quad $\triangleright$ Creates pipeline integrating stored embeddings for improved responses \\
        \hline
        
        \texttt{\textbf{ChatOpenAI}} \textit{(OpenAI LLM Interface)} \\
        \quad $\triangleright$ Connects to OpenAI models (GPT-4o Mini) for response generation \\
        \hline
        
        \texttt{\textbf{ChatAnthropic}} \textit{(Anthropic LLM Interface)} \\
        \quad $\triangleright$ Connects to Anthropic Claude models (Claude 3.5 Sonnet) for responses \\
        \hline
        
    \end{tabular}
    \end{adjustbox}
    \label{tab:langchain_components}
\end{table}

When a user query is submitted (e.g., a quantum programming task or challenge description), we use Maximum Marginal Relevance (MMR) search to retrieve the most relevant examples from the dataset. The retrieved examples are concatenated into a structured prompt and passed to the LLM for code generation. For a fair assessment of RAG’s impact, we generate outputs for each query both with and without retrieved context under identical conditions.

\section{Statistical Evaluation of the Dataset \label{sec4}}
A comprehensive analysis of the dataset was conducted to assess the distribution of instruction lengths, response lengths, response variations across feature groups, and their correlation with quantum functionalities. This section provides insights into the dataset's structure, response variability, and its implications for training LLMs.

\subsection{Response Lengths Across Feature Groups}
\begin{figure}[htpb]
    \centering
    \includegraphics[width=\linewidth]{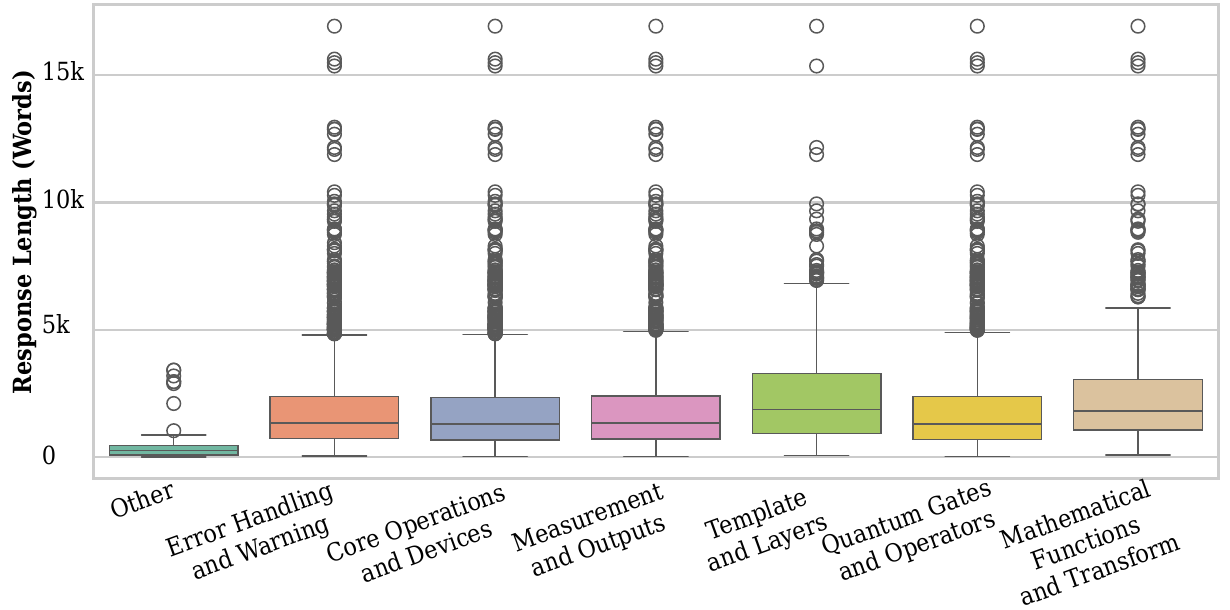}
    \caption{Box plot of response lengths across different feature groups. Each category represents a specific quantum computing feature set, with response lengths exhibiting variability and outliers, particularly in core operations, measurement, and mathematical functions.}
    \label{fig:resp_per_feature}
\end{figure}

Fig.~\ref{fig:resp_per_feature} provides a box plot representation of response lengths across different feature groups.

\begin{itemize}
    \item Response lengths remain relatively stable across feature groups, suggesting uniformity in generated outputs.
    \item Outliers appear in every feature category, indicating occasional long responses, particularly in core operations, measurement, and mathematical functions.
    \item The “Other” category has the shortest responses, likely due to simpler queries requiring minimal output.
\end{itemize}

\begin{figure*}[t]
    \centering
    \includegraphics[width=\linewidth]{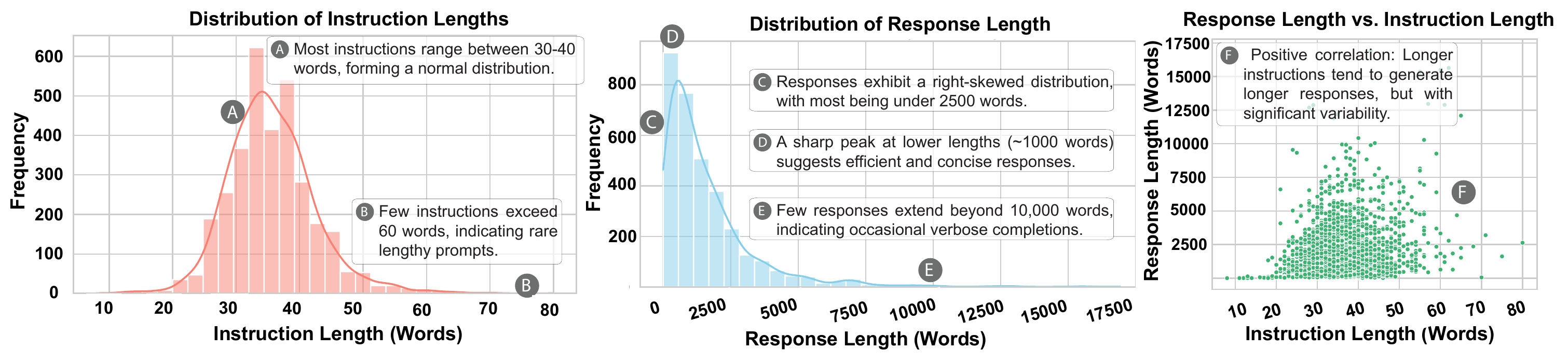}
    \caption{Distribution of instruction and response lengths with their correlation. The left plot shows the normal distribution of instruction lengths, the middle plot highlights the right-skewed distribution of response lengths, and the right scatter plot illustrates the relationship between instruction and response length.}
    \label{fig:instr_resp_distr}
\end{figure*}

\subsection{Instruction and Response Length Distributions}

Fig.~\ref{fig:instr_resp_distr} illustrates the distribution of instruction and response lengths, highlighting key patterns in the dataset. Instruction lengths, as shown in left plot, follow a normal distribution with a peak around 35 words. The majority of instructions fall within the 30–40 word range, while longer prompts exceeding 60 words are relatively rare. In contrast, response lengths exhibit a right-skewed distribution, as shown in the middle plot. Most responses are under 2500 words, with a distinct peak around 1000 words, though some responses exceed 10,000 words, indicating instances of verbose outputs. The instruction-response correlation, illustrated in the rightmost plot, reveals a positive correlation between instruction and response length. While longer instructions tend to produce longer responses, variability increases significantly beyond 50 words, highlighting the influence of instruction complexity on response verbosity.

\subsection{Feature Group Usage Across Quantum Domains}

\begin{figure}[htpb]
    \centering
    \includegraphics[width=\linewidth]{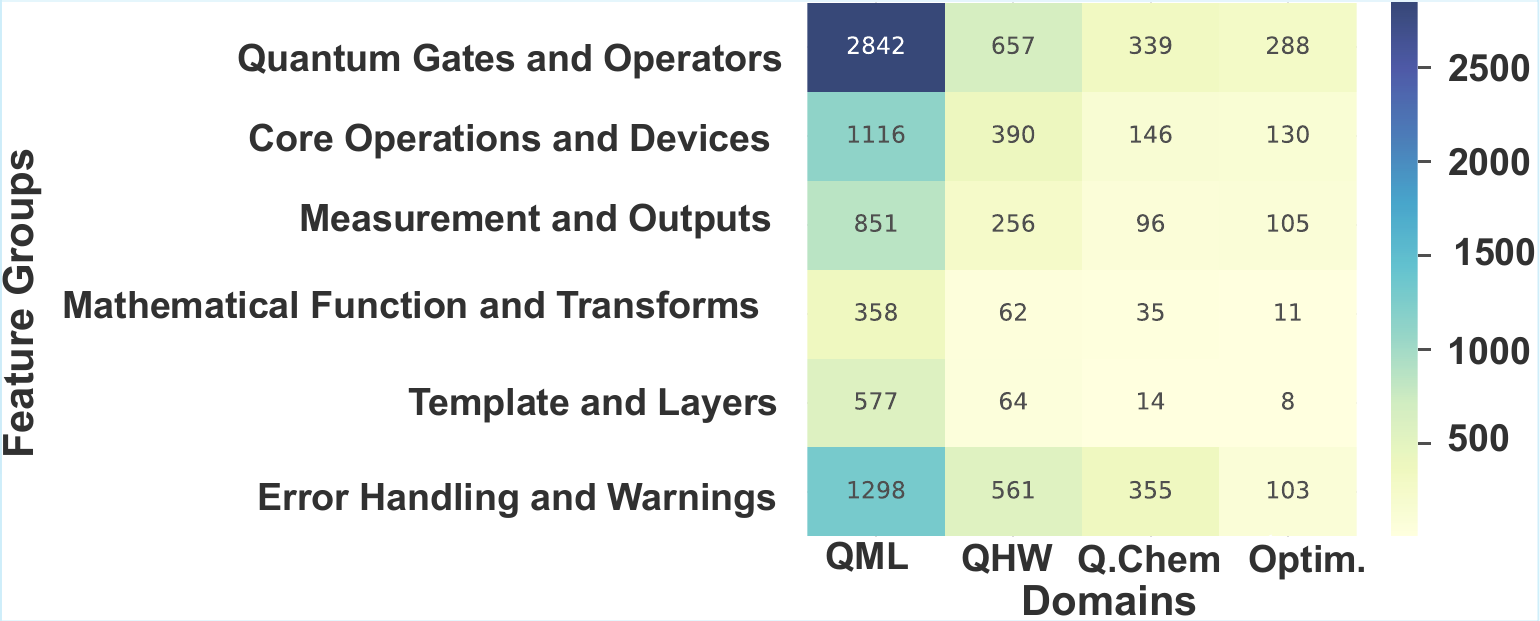}
    \caption{Feature group usage across domains: Heatmap illustrating how frequently each quantum feature group appears across the different domain categories (QML, Quantum Hardware, Quantum Chemistry and Optimization), with color intensity reflecting count.}
    \label{fig:domain_usage}
\end{figure}

The relationship between feature groups and quantum computing domains, such as machine learning, hardware, chemistry, and optimization, is illustrated in Fig.~\ref{fig:domain_usage}. Quantum Machine Learning relies predominantly on quantum gates and operators, alongside error handling techniques, reflecting its dependence on structured quantum transformations and noise mitigation. In contrast, Quantum Chemistry and Optimization exhibit relatively lower usage of mathematical functions, suggesting that these domains rely on alternative computational approaches rather than extensive arithmetic operations. Quantum Hardware applications, however, demonstrate a balanced distribution across various feature groups, indicating a broader, more integrated usage of quantum functionalities across different operational layers.

\subsection{Feature Category Association with Quantum Functions}

Quantum gates and operators exhibit the strongest associations with multiple features, particularly with commonly used functions such as qml.PauliZ, qml.RX, and qml.RY, which are integral to quantum circuit construction. Core operations and measurement functions also display high correlations, reflecting their frequent application in quantum circuit execution and result interpretation. Meanwhile, error handling and mathematical functions show relatively lower but still significant interactions with select quantum features, indicating their importance in specific computational scenarios rather than across all use cases.
\subsection{Insights and Implications}
Findings from Figures~\ref{fig:resp_per_feature}, \ref{fig:instr_resp_distr}, and \ref{fig:domain_usage} offer several insights:

\begin{itemize}
    \item \textbf{Balanced Instruction Lengths}: The normal distribution of instruction lengths ensures the dataset is well-structured for model training.
    \item \textbf{Response Length Variability}: Most responses are concise, but outliers indicate that certain quantum features demand more detailed explanations.
    \item \textbf{Feature-Specific Trends}: Quantum gates and measurement functions dominate dataset interactions, aligning with core PennyLane applications.
    \item \textbf{Domain-Specific Usage}: Different quantum fields rely on distinct feature sets, highlighting potential customization opportunities for specialized LLM training.
\end{itemize}

\section{Evaluation}
\label{sec:Evaluation}

To assess the effectiveness of our proposed dataset for quantum code generation, we designed a benchmark suite comprising 264 test cases, each tailored to evaluate various PennyLane functionalities and quantum programming concepts. 
The test cases (prompts) cover a broad spectrum of tasks, ranging from basic circuit construction to advanced hybrid algorithm design, encompassing state preparation, parameterized gate implementation, differentiation routines, and optimization workflows. Several tasks address realistic applications, including molecular energy estimation via the Variational Quantum Eigensolver (VQE) for systems such as $\mathrm{H_2}$ and $\mathrm{LiH}$, as well as hybrid quantum–classical models trained on toy datasets for classification and regression. An example of an advanced prompt is:
\begin{quote}
\textit{``Write a PennyLane program that implements a Variational Quantum Eigensolver to estimate the ground-state energy of the $\mathrm{H_2}$ molecule using the \texttt{default.qubit} device. Define the molecular Hamiltonian, construct a parameterized ansatz circuit, and optimize its parameters using gradient-based optimization until convergence.''}
\end{quote}

Each test was used as a prompt within an RAG pipeline and evaluated across both open-source and commercial LLMs under different context settings (100\%, 75\%, 50\%, and 0\%), where 100\% denotes full retrieved context and 0\% corresponds to no RAG. This setup was used to evaluate the effect of retrieval on code generation performance. The evaluation employed \texttt{pass@1} through \texttt{pass@5}, where each model generated five candidate solutions per test. The success rate (\%) was computed as $\texttt{pass@k}/264 \times 100$, where \texttt{pass@k} denotes the number of test cases, out of 264, for which at least one correct solution was obtained within the first \texttt{k} attempts. A generated solution was counted as correct if the notebook executed successfully and satisfied the corresponding test-case checks for the given inputs. The evaluation focused on whether the generated code correctly implemented the required task and produced the expected valid output for that test case. For tasks involving learning or optimization, a solution was considered correct if the requested workflow executed properly and returned the required type of result, regardless of the final performance value. In total, 1,320 notebooks were generated and executed per model (264 tests $\times$ 5 completions), providing a rigorous code-level functional evaluation. Table~\ref{tab:llm_models} lists the evaluated models.

\begin{table}[ht!]
    \centering
    \large
    \caption{LLMs Used in the RAG-based Evaluation Framework}
    \begin{adjustbox}{max width=\linewidth}
    \begin{tabular}{|l|}
        \hline
        \textbf{Large Language Models in RAG Pipeline} \\
        \hline\hline
        
        \textbf{Qwen2.5-7B-Instruct-Turbo} \textit{(Alibaba Cloud)} \\
        \quad $\bullet$ \textbf{Parameters:} 7B $|$ \textbf{Architecture:} Causal Decoder-Only Transformer \\
        \quad $\bullet$ \textbf{Usage:} Generates PennyLane code \\
        \hline
        
        \textbf{LLaMa 4 Maverick} \textit{(Meta)} \\
        \quad $\bullet$ \textbf{Parameters:} 17B $|$ \textbf{Architecture:} Causal Decoder-Only Transformer \\
        \quad $\bullet$ \textbf{Usage:} Code generation \\
        \hline
        
        \textbf{GPT-4o Mini} \textit{(OpenAI)} \\
        \quad $\bullet$ \textbf{Parameters:} Not disclosed $|$ \textbf{Architecture:} Causal Decoder-Only Transformer \\
        \quad $\bullet$ \textbf{Usage:} Evaluates and scores generated responses \\
        \hline
        
        \textbf{Claude 3.5 Sonnet} \textit{(Anthropic)} \\
        \quad $\bullet$ \textbf{Parameters:} Not disclosed $|$ \textbf{Architecture:} Causal Decoder-Only Transformer \\
        \quad $\bullet$ \textbf{Usage:} Alternative model for response generation \\
        \hline

        \textbf{Gemini 2.5 Flash} \textit{(Google DeepMind)} \\
        \quad $\bullet$ \textbf{Parameters:} Not disclosed $|$ \textbf{Architecture:} Causal Decoder-Only Transformer \\
        \quad $\bullet$ \textbf{Usage:} Rapid retrieval-augmented generation and code reasoning \\
        \hline

        \textbf{Claude Haiku} \textit{(Anthropic)} \\
        \quad $\bullet$ \textbf{Parameters:} Not disclosed $|$ \textbf{Architecture:} Causal Decoder-Only Transformer \\
        \quad $\bullet$ \textbf{Usage:} Lightweight LLM for fast inference and evaluation tasks \\
        \hline

        \textbf{GPT-5 Mini} \textit{(OpenAI)} \\
        \quad $\bullet$ \textbf{Parameters:} Not disclosed $|$ \textbf{Architecture:} Causal Decoder-Only Transformer \\
        \quad $\bullet$ \textbf{Usage:} Enhanced multi-turn reasoning and tool-augmented evaluation \\
        \hline
        
    \end{tabular}
    \end{adjustbox}
    \label{tab:llm_models}
\end{table}

\begin{table*}[h]
\centering
%\scriptsize
\caption{Evaluation of Qwen 2.5 7B, LLaMa 4 Maverick 17B, GPT-4o Mini, and Claude 3.5 Sonnet on PennyLane. Results for open-source models (Qwen and LLaMa) are highlighted in purple to distinguish their RAG-based performance gains from commercial models.}
\begin{adjustbox}{max width=\linewidth}
\begin{tabular}{ccccccccccccccccc}
%{l@{\hspace{8pt}}|@{\hspace{6pt}}cccc@{\hspace{6pt}}|@{\hspace{6pt}}cccc@{\hspace{6pt}}|@{\hspace{6pt}}cccc@{\hspace{6pt}}|@{\hspace{6pt}}cccc}
\toprule
\multirow{2}{*}{\textbf{Metric}} & 
\multicolumn{4}{c@{\hspace{6pt}}|}{\textbf{Qwen 2.5 7B}} & 
\multicolumn{4}{c@{\hspace{6pt}}|}{\textbf{LLaMa 4 Maverick}} &
\multicolumn{4}{c@{\hspace{6pt}}|}{\textbf{GPT-4o Mini}} & 
\multicolumn{4}{c}{\textbf{Claude 3.5 Sonnet}} \\
\cmidrule(lr){2-5} \cmidrule(lr){6-9} \cmidrule(l){10-13} \cmidrule(l){14-17}
& \textbf{Full} & \textbf{75\%} & \textbf{50\%} & \textbf{0\%} & 
\textbf{Full} & \textbf{75\%} & \textbf{50\%} & \textbf{0\%} & 
\textbf{Full} & \textbf{75\%} & \textbf{50\%} & \textbf{0\%} & 
\textbf{Full} & \textbf{75\%} & \textbf{50\%} & \textbf{0\%} \\
\midrule
%\textbf{Total Notebooks} & \multicolumn{4}{c|}{\textbf{1320}} & \multicolumn{4}{c|}{\textbf{1320}} & \multicolumn{4}{c|}{\textbf{1320}} & \multicolumn{4}{c}{\textbf{1320}} \\
%\midrule
%\textbf{Task Success} & \textcolor{purple}{\textbf{35.4\%}} & \textcolor{purple}{\textbf{37.5\%}} & \textcolor{purple}{\textbf{36.4\%}} & 7.2\% & \textcolor{purple}{\textbf{68.3\%}} & \textcolor{purple}{\textbf{66.3\%}} & \textcolor{purple}{\textbf{67.8\%}} & 60.5\% & 53.9\% & 59.1\% & 57.2\% & 47.3\% & 68.1\% & 69.0\% & 64.8\% & 75.8\% \\
%\textbf{Working Notebooks} & \textcolor{purple}{\textbf{467}} & \textcolor{purple}{\textbf{495}} & \textcolor{purple}{\textbf{480}} & 95 & \textcolor{purple}{\textbf{901}} & \textcolor{purple}{\textbf{875}} & \textcolor{purple}{\textbf{895}} & 799 & 712 & 780 & 755 & 625 & 899 & 911 & 856 & 1001 \\
%\textbf{Success Rate (\%)} & \textcolor{purple}{\textbf{41.66}} & \textcolor{purple}{\textbf{45.07}} & \textcolor{purple}{\textbf{43.18}} & 8.71 & \textcolor{purple}{\textbf{84.84}} & \textcolor{purple}{\textbf{84.84}} & \textcolor{purple}{\textbf{82.17}} & 78.78 & 80.30 & 84.84 & 81.81 & 85.60 & 89.01 & 92.44 & 87.50 & 95.07 \\
%\midrule
\textbf{Pass@1} & \textcolor{purple}{\textbf{33.0\%}} & \textcolor{purple}{\textbf{37.9\%}} & \textcolor{purple}{\textbf{33.3\%}} & 7.2\% & \textcolor{purple}{\textbf{67.8\%}} & \textcolor{purple}{\textbf{64.8\%}} & \textcolor{purple}{\textbf{67.8\%}} & 56.8\% & 54.5\% & 59.8\% & 56.4\% & 50.8\% & 70.8\% & 68.9\% & 64.4\% & 73.5\% \\
\textbf{Pass@2} & \textcolor{purple}{\textbf{37.5\%}} & \textcolor{purple}{\textbf{41.3\%}} & \textcolor{purple}{\textbf{39.0\%}} & 8.0\% & \textcolor{purple}{\textbf{75.0\%}} & \textcolor{purple}{\textbf{76.5\%}} & \textcolor{purple}{\textbf{75.4\%}} & 67.4\% & 64.8\% & 71.6\% & 66.3\% & 66.3\% & 79.5\% & 82.6\% & 76.9\% & 87.1\% \\
\textbf{Pass@3} & \textcolor{purple}{\textbf{39.4\%}} & \textcolor{purple}{\textbf{42.8\%}} & \textcolor{purple}{\textbf{41.3\%}} & 8.0\% & \textcolor{purple}{\textbf{80.3\%}} & \textcolor{purple}{\textbf{81.1\%}} & \textcolor{purple}{\textbf{79.2\%}} & 72.3\% & 72.0\% & 78.4\% & 73.9\% & 75.4\% & 84.5\% & 87.9\% & 81.1\% & 90.2\% \\
\textbf{Pass@4} & \textcolor{purple}{\textbf{41.3\%}} & \textcolor{purple}{\textbf{44.3\%}} & \textcolor{purple}{\textbf{41.7\%}} & 8.3\% & \textcolor{purple}{\textbf{81.8\%}} & \textcolor{purple}{\textbf{82.2\%}} & \textcolor{purple}{\textbf{80.3\%}} & 76.5\% & 76.5\% & 82.9\% & 79.2\% & 81.4\% & 87.1\% & 91.7\% & 87.1\% & 93.2\% \\
\textbf{Pass@5} & \textcolor{purple}{\textbf{41.7\%}} & \textcolor{purple}{\textbf{45.1\%}} & \textcolor{purple}{\textbf{43.2\%}} & 8.7\% & \textcolor{purple}{\textbf{84.8\%}} & \textcolor{purple}{\textbf{84.8\%}} & \textcolor{purple}{\textbf{82.2\%}} & 78.8\% & 80.3\% & 84.8\% & 81.8\% & 85.6\% & 89.0\% & 92.4\% & 87.5\% & 95.1\% \\
%\textbf{Pass@1} & \textcolor{purple}{\textbf{87}} & \textcolor{purple}{\textbf{100}} & \textcolor{purple}{\textbf{88}} & 19 & \textcolor{purple}{\textbf{179}} & \textcolor{purple}{\textbf{171}} & \textcolor{purple}{\textbf{179}} & 150 & 144 & 158 & 149 & 134 & 187 & 182 & 170 & 194 \\
%\textbf{Pass@2} & \textcolor{purple}{\textbf{99}} & \textcolor{purple}{\textbf{109}} & \textcolor{purple}{\textbf{103}} & 21 & \textcolor{purple}{\textbf{198}} & \textcolor{purple}{\textbf{202}} & \textcolor{purple}{\textbf{199}} & 178 & 171 & 189 & 175 & 175 & 210 & 218 & 203 & 230 \\
%\textbf{Pass@3} & \textcolor{purple}{\textbf{104}} & \textcolor{purple}{\textbf{113}} & \textcolor{purple}{\textbf{109}} & 21 & \textcolor{purple}{\textbf{212}} & \textcolor{purple}{\textbf{214}} & \textcolor{purple}{\textbf{209}} & 191 & 190 & 207 & 195 & 199 & 223 & 232 & 214 & 238 \\
%\textbf{Pass@4} & \textcolor{purple}{\textbf{109}} & \textcolor{purple}{\textbf{117}} & \textcolor{purple}{\textbf{110}} & 22 & \textcolor{purple}{\textbf{216}} & \textcolor{purple}{\textbf{217}} & \textcolor{purple}{\textbf{212}} & 202 & 202 & 219 & 209 & 215 & 230 & 242 & 230 & 246 \\
%\textbf{Pass@5} & \textcolor{purple}{\textbf{110}} & \textcolor{purple}{\textbf{119}} & \textcolor{purple}{\textbf{114}} & 23 & \textcolor{purple}{\textbf{224}} & \textcolor{purple}{\textbf{224}} & \textcolor{purple}{\textbf{217}} & 208 & 212 & 224 & 216 & 226 & 235 & 244 & 231 & 251 \\
\bottomrule
\end{tabular}
\end{adjustbox}
\label{tab:all-models}
\end{table*}

\begin{figure}[htpb]
    \centering
    \includegraphics[width=\linewidth]{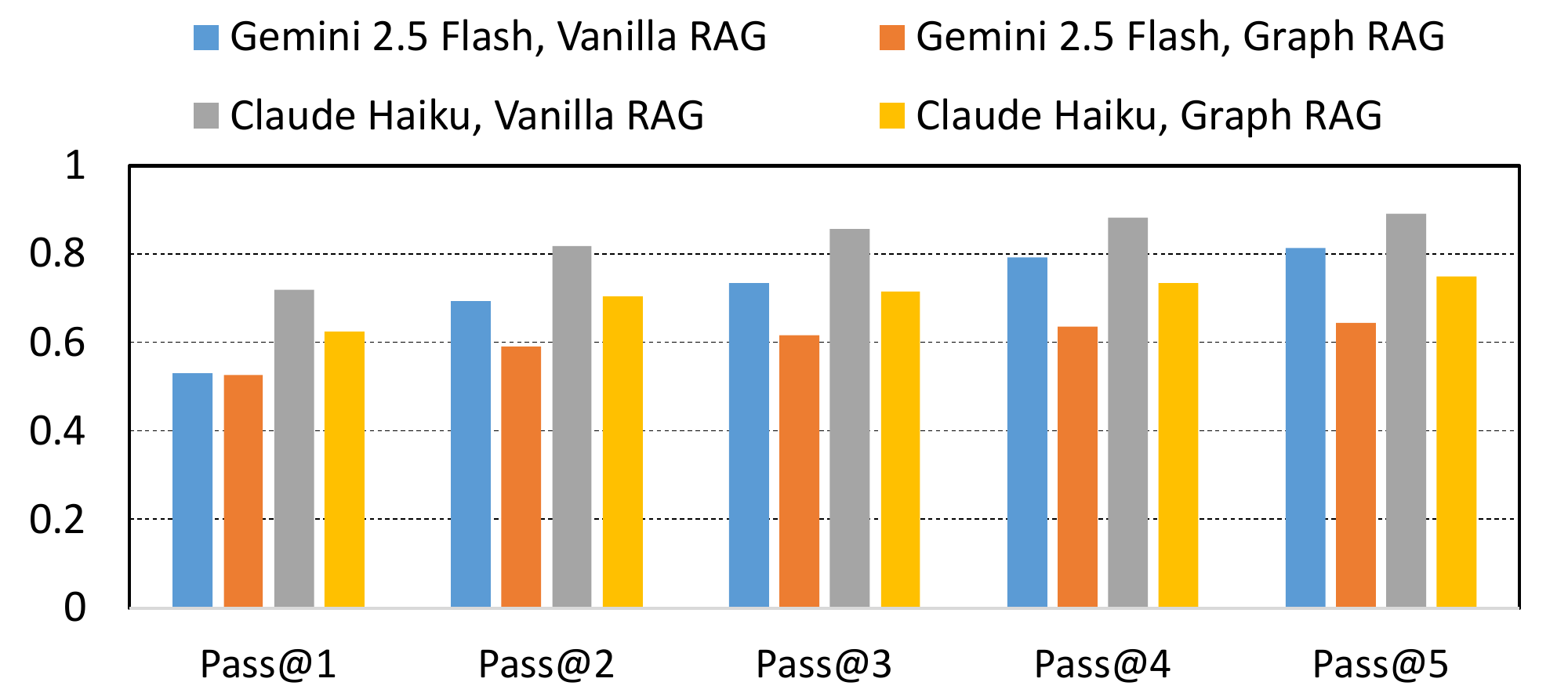}
    \caption{Ablation Study: Vanilla RAG vs. Graph RAG evaluated on Gemini 2.5 Flash and Claude Haiku.}
    \label{fig:graphrag_ablation}
\end{figure}

\paragraph{Analysis: Vanilla RAG vs. GraphRAG.}
Figure~\ref{fig:graphrag_ablation} compares \emph{vanilla} retrieval-augmented generation (RAG) against a GraphRAG variant on our PennyLane benchmark, using two representative commercial models (Gemini~2.5~Flash and Claude~Haiku).
Across all \texttt{pass@k} metrics, vanilla RAG consistently outperforms GraphRAG for both models, indicating that the added graph-structuring step does not translate into better executable-code success in this setting.
Overall, these results suggest that, for our dataset and tasks, the primary gains come from retrieving a small set of \emph{directly relevant} PennyLane exemplars rather than from explicitly modeling inter-document relations via a graph.

Since our benchmark emphasizes functional correctness of generated notebooks, where small syntactic or API-level deviations can cause execution failures, the precision of the retrieved snippets is often more important than broader relational coverage.
Graph construction can also impose additional routing/selection constraints (e.g., traversing nodes that are only indirectly related), which may reduce the density of immediately usable code patterns in the final prompt compared to standard similarity-based retrieval.
Given that vanilla RAG consistently outperforms GraphRAG in this ablation study, we adopt vanilla RAG as the default retrieval strategy for the remainder of the evaluations.

\paragraph{Open-Source Models.}
 Given that the dataset was primarily developed to support open-source development, our core evaluation focused on Qwen 7B and LLaMa 4. As shown in Table~\ref{tab:all-models}, these models showed clear improvements when used with our dataset. Notably, RAG with 75\% context coverage often outperformed the full-context setting. This suggests that a moderate amount of relevant context may reduce noise and improve grounding, while excessive retrieval might introduce redundancy or irrelevant tokens.
\paragraph{Commercial Models.}
 For comparison, we also tested GPT-4o Mini, Claude 3.5 Sonnet, Gemini 2.5 Flash, Claude Haiku, and GPT-5 Mini. As shown in Tables~\ref{tab:all-models} and~\ref{tab:new-models-75-0}, these commercial models demonstrated high performance even without RAG. However, the added context from our dataset provided limited benefit. This may be attributed to their broad pretraining, which likely includes similar material (e.g., PennyLane documentation or equivalent quantum code examples), reducing the marginal gain from external retrieval.

 \begin{table}[h]
\centering
\caption{Evaluation of Gemini 2.5 Flash, Claude Haiku, and GPT-5 Mini on PennyLane under two RAG coverage settings (75\% and 0\%).}
\begin{adjustbox}{max width=\linewidth}
\begin{tabular}{ccccccc}%{l@{\hspace{8pt}}|@{\hspace{6pt}}cc@{\hspace{6pt}}|@{\hspace{6pt}}cc@{\hspace{6pt}}|@{\hspace{6pt}}cc@{\hspace{6pt}}}
\toprule
\multirow{2}{*}{\textbf{Metric}} &
\multicolumn{2}{c@{\hspace{6pt}}|}{\textbf{Gemini 2.5 Flash}} &
\multicolumn{2}{c@{\hspace{6pt}}|}{\textbf{Claude Haiku}} &
\multicolumn{2}{c@{\hspace{6pt}}}{\textbf{GPT-5 Mini}} \\
\cmidrule(lr){2-3} \cmidrule(lr){4-5} \cmidrule(lr){6-7}
& \textbf{75\%} & \textbf{0\%} & \textbf{75\%} & \textbf{0\%} & \textbf{75\%} & \textbf{0\%} \\
\midrule
% \textbf{Total Notebooks}                & \textbf{1320} & \textbf{1320} & \textbf{1320} & \textbf{1320} & \textbf{1320} & \textbf{--} \\
% \textbf{Working Notebooks}              & \textbf{721}  & \textbf{671}  & \textbf{945}  & \textbf{862}  & \textbf{973}  & \textbf{--} \\
% \textbf{Notebook Success Rate (\%)}     & \textbf{54.62} & \textbf{50.83} & \textbf{71.59} & \textbf{65.30} & \textbf{73.71} & \textbf{--} \\
% \midrule
%\textbf{Total Tasks}                    & \textbf{264}  & \textbf{264}  & \textbf{264}  & \textbf{264}  & \textbf{264}  & \textbf{264} \\
%\textbf{Task Success (\%)}     & \textbf{81.44} & \textbf{75.00} & \textbf{89.02} & \textbf{89.77} & \textbf{91.29} & \textbf{92.04} \\
%\midrule
\textbf{Pass@1} & 53.0\% & 55.3\% & 72.0\% & 67.4\% & 73.9\% & 74.6\% \\
\textbf{Pass@2} & 69.3\% & 64.8\% & 81.8\% & 78.8\% & 85.6\% & 85.2\% \\
\textbf{Pass@3} & 73.5\% & 68.9\% & 85.6\% & 84.1\% & 89.8\% & 88.3\% \\
\textbf{Pass@4} & 79.2\% & 72.0\% & 88.3\% & 87.9\% & 90.5\% & 89.8\% \\
\textbf{Pass@5} & 81.4\% & 75.0\% & 89.0\% & 89.8\% & 91.3\% & 92.0\% \\
%\textbf{Pass@1} & \textbf{53.0\%} & \textbf{55.3\%} & \textbf{72.0\%} & \textbf{67.4\%} & \textbf{73.9\%} & \textbf{74.6\%} \\
%\textbf{Pass@2} & \textbf{69.3\%} & \textbf{64.8\%} & \textbf{81.8\%} & \textbf{78.8\%} & \textbf{85.6\%} & \textbf{85.2\%} \\
%\textbf{Pass@3} & \textbf{73.5\%} & \textbf{68.9\%} & \textbf{85.6\%} & \textbf{84.1\%} & \textbf{89.8\%} & \textbf{88.3\%} \\
%\textbf{Pass@4} & \textbf{79.2\%} & \textbf{72.0\%} & \textbf{88.3\%} & \textbf{87.9\%} & \textbf{90.5\%} & \textbf{89.8\%} \\
%\textbf{Pass@5} & \textbf{81.4\%} & \textbf{75.0\%} & \textbf{89.0\%} & \textbf{89.8\%} & \textbf{91.3\%} & \textbf{92.0\%} \\
%\textbf{Pass@1}                         & \textbf{140}  & \textbf{146}  & \textbf{190}  & \textbf{178}  & \textbf{195}  & \textbf{197} \\
%\textbf{Pass@2}                         & \textbf{183}  & \textbf{171}  & \textbf{216}  & \textbf{208}  & \textbf{226}  & \textbf{225} \\
%\textbf{Pass@3}                         & \textbf{194}  & \textbf{182}  & \textbf{226}  & \textbf{222}  & \textbf{237}  & \textbf{233} \\
%\textbf{Pass@4}                         & \textbf{209}  & \textbf{190}  & \textbf{233}  & \textbf{232}  & \textbf{239}  & \textbf{237} \\
%\textbf{Pass@5}                         & \textbf{215}  & \textbf{198}  & \textbf{235}  & \textbf{237}  & \textbf{241}  & \textbf{243} \\
\bottomrule
\end{tabular}
\end{adjustbox}
\label{tab:new-models-75-0}
\end{table}

\paragraph{Discussion.}
 The evaluation reveals important interactions between dataset design, model capacity, and retrieval strategies. Open-source models such as Qwen 7B and LLaMa 4 showed notable improvements when augmented with \textit{PennyLang}, confirming that domain-specific RAG can effectively compensate for limited pretraining in quantum programming. Interestingly, the 75\% context setting consistently outperformed the full-context retrieval, indicating that focused and moderately sized context windows improve model grounding by avoiding redundancy and context saturation.

In contrast, commercial models like GPT-4o Mini, Claude 3.5 Sonnet, Gemini 2.5 Flash, Claude Haiku, and GPT-5 Mini achieved high performance across all settings, with only marginal benefits from added retrieval. This behavior is likely due to their extensive pretraining on diverse codebases, which may already include material similar to what our dataset provides. Consequently, the impact of external retrieval is diminished and can even introduce noise if not precisely targeted.
These findings emphasize the importance of retrieval quality over quantity in RAG-based code generation. The consistent advantage of partial over full context suggests that carefully curated, relevant examples are more beneficial than exhaustive inclusion. 

Finally, by evaluating 1,320 generated notebooks per model, we ensure that our conclusions are grounded in executable and functionally correct outputs, a crucial criterion in the domain of quantum programming. These results underscore the utility of \textit{PennyLang} in enhancing open-source LLM performance while providing insights into optimal RAG configurations for technical code generation tasks.

\section{Conclusion \label{sec5}}
In this paper, we introduce \textit{PennyLang}, a high-quality instruction–response dataset tailored for PennyLane-based quantum code generation. We present our data collection and preprocessing methodology, drawing from GitHub repositories, textbooks, and official documentation to curate 3,347 diverse and well-structured code samples. To evaluate its practical utility, we develop a RAG evaluation framework and benchmark seven state-of-the-art LLMs (both open-source and commercial models) under both RAG and non-RAG settings. Our findings show that \textit{PennyLang}, when paired with RAG, substantially boosts performance for open-source models, while proprietary models exhibit minimal or negative gains with naive full-context retrieval, likely because their pre-training already covers overlapping PennyLane-related knowledge. This further highlights the importance of calibrated retrieval budgets, as moderate settings (e.g., 75\%) often outperform full context by balancing signal and noise. We release both the \textit{PennyLang} dataset and the evaluation framework to foster reproducibility and encourage further research in LLM-assisted quantum development. Future directions include training lightweight instruction-tuned models on this dataset, improving the filtering pipeline for continuously collecting high-quality new code snippets into the database, optimizing retrieval strategies across different model types, and extending to cross-framework generalization between PennyLane, Qiskit, and Cirq.

\section*{Data Availability}
The dataset proposed in this paper is publicly available on GitHub at \url{https://github.com/eBrain4Everyone/Pennylang}.
\section*{Acknowledgment}
 This work was supported in part by the NYUAD Center for Quantum and Topological Systems (CQTS), funded by Tamkeen under the NYUAD Research Institute grant CG008, and Center for CyberSecurity (CCS), funded by Tamkeen under the NYUAD Research Institute Award G1104. This research was carried out on the High Performance Computing resources at New York University Abu Dhabi.

\bstctlcite{IEEEexample:BSTcontrol}
\begin{spacing}{0.92}
    \bibliographystyle{IEEEtran}
\bibliography{cite}

@IEEEtranBSTCTL{IEEEexample:BSTcontrol,
CTLdash_repeated_names = "no"
}

@Article{feynman1982simulating,
author={Feynman, Richard P.},
title={Simulating physics with computers},
journal={International Journal of Theoretical Physics},
year={1982},
month={Jun},
day={01},
volume={21},
number={6},
pages={467-488},
issn={1572-9575},
doi11={10.1007/BF02650179},
}

@book{ganguly2021quantum,
  title={Quantum machine learning: an applied approach},
  author={Ganguly, Santanu},
  year={2021},
  publisher={Springer}
}

@article{RAG,
  title={Retrieval-augmented generation for knowledge-intensive nlp tasks},
  author2={Lewis, Patrick and Perez, Ethan and Piktus, Aleksandra and Petroni, Fabio and Karpukhin, Vladimir and Goyal, Naman and K{\"u}ttler, Heinrich and Lewis, Mike and Yih, Wen-tau and Rockt{\"a}schel, Tim and others},
  author={Lewis, Patrick and others},
  journal={Advances in neural information processing systems},
  volume2={33},
  pages2={9459--9474},
  year={2020}
}

@inproceedings{pathak2024resource,
  title={Resource Allocation Optimization in 5G Networks Using Variational Quantum Regressor},
  author={Pathak, Param and Oad, Vidhi and Prajapati, Aditya and Innan, Nouhaila},
  booktitle={2024 International Conference on Quantum Communications, Networking, and Computing (QCNC)},
  pages={101--105},
  year={2024},
  organization={IEEE}
}

@INPROCEEDINGS{10651123,
  author={Innan, Nouhaila and Khan, Muhammad Al-Zafar and Marchisio, Alberto and Shafique, Muhammad and Bennai, Mohamed},
  booktitle={2024 International Joint Conference on Neural Networks (IJCNN)}, 
  title={FedQNN: Federated Learning using Quantum Neural Networks}, 
  year={2024},
  volume={},
  number={},
  pages2={1-9},
  doi11={10.1109/IJCNN60899.2024.10651123}}

@article{innan2024financial,
  title={Financial fraud detection using quantum graph neural networks},
  author={Innan, Nouhaila and Sawaika, Abhishek and Dhor, Ashim and Dutta, Siddhant and Thota, Sairupa and Gokal, Husayn and Patel, Nandan and Khan, Muhammad Al-Zafar and Theodonis, Ioannis and Bennai, Mohamed},
  journal={Quantum Machine Intelligence},
  volume={6},
  number={1},
  pages={7},
  year={2024},
  publisher={Springer}
}

@book{combarro2023practical,
  title={A practical guide to quantum machine learning and quantum optimization: Hands-on approach to modern quantum algorithms},
  author2={Combarro, Elias F and Gonz{\'a}lez-Castillo, Samuel and Di Meglio, Alberto},
  author={Combarro, Elias F and others},
  year={2023},
  publisher={Packt Publishing Ltd}
}

@article{innan2024quantum1,
  title={Quantum computing for electronic structure analysis: Ground state energy and molecular properties calculations},
  author={Innan, Nouhaila and Khan, Muhammad Al-Zafar and Bennai, Mohamed},
  journal={Materials Today Communications},
  volume2={38},
  pages2={107760},
  year={2024},
  publisher={Elsevier}
}

@article{preskill2018nisq,
   title={Quantum Computing in the NISQ era and beyond},
   volume={2},
   ISSN={2521-327X},
   url1={http://dx.doi1.org/10.22331/q-2018-08-06-79},
   doi1={10.22331/q-2018-08-06-79},
   journal={Quantum},
   publisher={Verein zur Forderung des Open Access Publizierens in den Quantenwissenschaften},
   author={Preskill, John},
   year={2018},
   month=aug, pages={79} }

@Article{arute2019quantum,
author={Arute, Frank
and Arya, Kunal
and Babbush, Ryan
and others},
title={Quantum supremacy using a programmable superconducting processor},
journal={Nature},
year={2019},
month={Oct},
day={01},
volume={574},
number={7779},
pages={505-510},
issn={1476-4687},
doi1={10.1038/s41586-019-1666-5},
url1={https://doi1.org/10.1038/s41586-019-1666-5}
}

@article{montanaro2016quantum,
   title={Quantum algorithms: an overview},
   volume={2},
   ISSN={2056-6387},
   url1={http://dx.doi1.org/10.1038/npjqi.2015.23},
   doi1={10.1038/npjqi.2015.23},
   number={1},
   journal={npj Quantum Information},
   publisher={Springer Science and Business Media LLC},
   author={Montanaro, Ashley},
   year={2016},
   month=jan }

@article{moll2018quantum,
   title={Quantum optimization using variational algorithms on near-term quantum devices},
   volume2={3},
   ISSN2={2058-9565},
   url1={http://dx.doi1.org/10.1088/2058-9565/aab822},
   doi1={10.1088/2058-9565/aab822},
   number2={3},
   journal={Quantum Science and Technology},
   publisher2={IOP Publishing},
   author2={Moll, Nikolaj and Barkoutsos, Panagiotis and Bishop, Lev S and Chow, Jerry M and Cross, Andrew and Egger, Daniel J and Filipp, Stefan and Fuhrer, Andreas and Gambetta, Jay M and Ganzhorn, Marc and Kandala, Abhinav and Mezzacapo, Antonio and Müller, Peter and Riess, Walter and Salis, Gian and Smolin, John and Tavernelli, Ivano and Temme, Kristan},
   author={Moll, Nikolaj and others},
   year={2018},
   month2=jun, pages2={030503} }

@article{Wang_2024,
   title={A comprehensive review of quantum machine learning: from NISQ to fault tolerance},
   volume={87},
   ISSN={1361-6633},
   url1={http://dx.doi1.org/10.1088/1361-6633/ad7f69},
   doi1={10.1088/1361-6633/ad7f69},
   number={11},
   journal={Reports on Progress in Physics},
   publisher={IOP Publishing},
   author={Wang, Yunfei and Liu, Junyu},
   year={2024},
   month=oct, pages={116402} }

@article{bergholm2018pennylane,
  title={Pennylane: Automatic differentiation of hybrid quantum-classical computations},
  author2={Bergholm, Ville and Izaac, Josh and Schuld, Maria and Gogolin, Christian and Ahmed, Shahnawaz and Ajith, Vishnu and Alam, M Sohaib and Alonso-Linaje, Guillermo and AkashNarayanan, Bharath and Asadi, Ali and others},
  author={Bergholm, Ville and others},
  journal={arXiv:1811.04968},
  year={2018}
}

@article{aleksandrowicz2019qiskit,
  author = {Aleksandrowicz, G. and Alexander, T. and Barkoutsos, P. and others},
  title = {Qiskit: an open-source framework for quantum computing},
  year = {2019},
  doi1 = {10.5281/zenodo.2562110}
}

@inproceedings{qiskitAssistant,
  title={Qiskit code assistant: Training llms for generating quantum computing code},
  author2={Dupuis, Nicolas and Buratti, Luca and Vishwakarma, Sanjay and Forrat, Aitana Viudes and Kremer, David and Faro, Ismael and Puri, Ruchir and Cruz-Benito, Juan},
  author={Dupuis, Nicolas and others},
  booktitle={2024 IEEE LLM Aided Design Workshop (LAD)},
  pages2={1--4},
  year={2024},
  organization2={IEEE}
}

@article{brown2020language,
  title={Language models are few-shot learners},
  author2={Brown, Tom and Mann, Benjamin and Ryder, Nick and Subbiah, Melanie and Kaplan, Jared D and Dhariwal, Prafulla and Neelakantan, Arvind and Shyam, Pranav and Sastry, Girish and Askell, Amanda and others},
  author={Brown, Tom and others},
  journal={Advances in neural information processing systems},
  volume2={33},
  pages2={1877--1901},
  year={2020}
}

@article{chen2021evaluating,
  author = {Chen, Mark and Tworek, Jerry and Jun, Heewoo and Yuan, Qiming and others},
  title = {Evaluating Large Language Models Trained on Code},
  journal = {arXiv:2107.03374},
  volume2 = {arXiv:2107.03374},
  year = {2021},
  url1 = {https://arxiv.org/abs/2107.03374}
}

@article{li2022competition,
  author = {Li, Yujia and Zhou, David R and Kohl, Nadine and Wu, Jamie and others},
  title = {Competition-Level Code Generation with AlphaCode},
  journal = {arXiv:2203.07814},
  volume2 = {arXiv:2203.07814},
  year = {2022},
  url1 = {https://arxiv.org/abs/2203.07814}
}

@misc{codex2021,
  author = {{OpenAI}},
  title = {{OpenAI Codex: Programming with Natural Language}},
  year = {2021},
  journal = {OpenAI Blog},
  howpublished = {\url{https://openai.com/index/openai-codex/}},
}

@article{codeLlama,
  title={Code llama: Open foundation models for code},
  author2={Roziere, Baptiste and Gehring, Jonas and Gloeckle, Fabian and Sootla, Sten and Gat, Itai and Tan, Xiaoqing Ellen and Adi, Yossi and Liu, Jingyu and Sauvestre, Romain and Remez, Tal and others},
  author={Roziere, Baptiste and others},
  journal={arXiv:2308.12950},
  year={2023}
}

@article{nijkamp2023codegen,
  title={Codegen: An open large language model for code with multi-turn program synthesis},
  author2={Nijkamp, Erik and Pang, Bo and Hayashi, Hiroaki and Tu, Lifu and Wang, Huan and Zhou, Yingbo and Savarese, Silvio and Xiong, Caiming},
  author={Nijkamp, Erik and others},
  journal={arXiv:2203.13474},
  year={2022}
}

@article{fried2022incoder,
  title={Incoder: A generative model for code infilling and synthesis},
  author2={Fried, Daniel and Aghajanyan, Armen and Lin, Jessy and Wang, Sida and Wallace, Eric and Shi, Freda and Zhong, Ruiqi and Yih, Wen-tau and Zettlemoyer, Luke and Lewis, Mike},
  author={Fried, Daniel and others},
  journal={arXiv:2204.05999},
  year={2022}
}

@book{schuld2018supervised,
  author = {Schuld, Maria and Petruccione, Francesco},
  title = {{Supervised Learning with Quantum Computers}},
  publisher = {Springer},
  year = {2018},
  edition = {1st},
  isbn = {978-3-319-96424-9},
  doi1 = {https://doi1.org/10.1007/978-3-319-96424-9}
}

@inproceedings{Ren2024,
author = {Ren, Xiangyu and Zhang, Tianyu and Xu, Xiong and Zheng, Yi-Cong and Zhang, Shengyu},
title = {Invited: Leveraging Machine Learning for Quantum Compilation Optimization},
year = {2024},
isbn2 = {9798400706011},
publisher2 = {Association for Computing Machinery},
booktitle = {DAC},
address2 = {New York, NY, USA},
url1 = {https://doi1.org/10.1145/3649329.3663510},
doi1 = {10.1145/3649329.3663510},
articleno2 = {360},
numpages2 = {4},
location2 = {San Francisco, CA, USA},
series2 = {DAC '24}
}

@inproceedings{feng2020codebert,
  title={Codebert: A pre-trained model for programming and natural languages},
  author2={Feng, Zhangyin and Guo, Daya and Tang, Duyu and Duan, Nan and Feng, Xiaocheng and Gong, Ming and Shou, Linjun and Qin, Bing and Liu, Ting and Jiang, Daxin and others},
  author={Feng, Zhangyin and others},
  booktitle={Findings of the association for computational linguistics: EMNLP 2020},
  pages2={1536--1547},
  year={2020}
}

@article{austin2021mbpp,
  title={Program synthesis with large language models},
  author2={Austin, Jacob and Odena, Augustus and Nye, Maxwell and Bosma, Maarten and Michalewski, Henryk and Dohan, David and Jiang, Ellen and Cai, Carrie and Terry, Michael and Le, Quoc and others},
  author={Austin, Jacob and others},
  journal={arXiv:2108.07732},
  year={2021}
}

@article{lu2021codexglue,
  title={Codexglue: A machine learning benchmark dataset for code understanding and generation},
  author2={Lu, Shuai and Guo, Daya and Ren, Shuo and Huang, Junjie and Svyatkovskiy, Alexey and Blanco, Ambrosio and Clement, Colin and Drain, Dawn and Jiang, Daxin and Tang, Duyu and others},
  author={Lu, Shuai and others},
  journal={arXiv:2102.04664},
  year={2021}
}

@article{li2023adaptingpretrainedlanguagemodels,
  title={Adapting pre-trained language models for quantum natural language processing},
  author2={Li, Qiuchi and Wang, Benyou and Zhu, Yudong and Lioma, Christina and Liu, Qun},
  author={Li, Qiuchi and others},
  journal={arXiv:2302.13812},
  year={2023}
}

@article{mitQuantumData,
  title={QCircuitNet: A large-scale hierarchical dataset for quantum algorithm design},
  author={Yang, Rui and Gu, Yuntian and Wang, Ziruo and Liang, Yitao and Li, Tongyang},
  journal={arXiv:2410.07961},
  year={2024}
}

@article{pennylaneaiGitHub,
  author = {{PennyLane Developer}},
  title = {PennyLaneAI GitHub Repository},
  year = {2023},
  note = {Available at \url{https://github.com/PennyLaneAI}}
}

@article{pennylaneDocs,
  author = {{PennyLane Developers}},
  title = {Official PennyLane Documentation},
  year = {2023},
  note = {Available at \url{https://docs.pennylane.ai/en/stable/}}
}

@article{van2001pep8,
  author = {Van Rossum, Guido and Warsaw, Barry and Coghlan, Nick},
  title = {PEP 8 – Style Guide for Python Code},
  year = {2001},
  note = {Available at \url{https://peps.python.org/pep-0008/}}
}

@inproceedings{huggingfaceTransformers,
    title = "Transformers: State-of-the-Art Natural Language Processing",
    author = "Wolf, Thomas  and
      Debut, Lysandre  and
      Sanh, Victor  and
      others",
    editor = "Liu, Qun  and
      Schlangen, David",
    booktitle = "Proceedings of the 2020 Conference on Empirical Methods in Natural Language Processing: System Demonstrations",
    month = oct,
    year = "2020",
    address = "Online",
    publisher = "Association for Computational Linguistics",
    url1 = "https://aclanthology.org/2020.emnlp-demos.6/",
    doi1 = "10.18653/v1/2020.emnlp-demos.6",
    pages = "38--45",
}

@article{transformerVaswani,
  title={Attention is all you need},
  author2={Vaswani, Ashish and Shazeer, Noam and Parmar, Niki and Uszkoreit, Jakob and Jones, Llion and Gomez, Aidan N and Kaiser, {\L}ukasz and Polosukhin, Illia},
  author={Vaswani, Ashish and others},
  journal={Advances in neural information processing systems},
  volume2={30},
  year={2017}
}

@article{attentionMechanism,
  title={Neural machine translation by jointly learning to align and translate},
  author={Bahdanau, Dzmitry and Cho, Kyunghyun and Bengio, Yoshua},
  journal={arXiv:1409.0473},
  year={2014}
}

@misc{qiskit_textbook,
  author = {Qiskit Community},
  title = {The Qiskit Textbook},
  year = {2023},
  url1 = {https://github.com/Qiskit/textbook}
}

@misc{ibm_quantum_challenge,
  author = {{IBM Quantum}},
  title = {IBM Quantum Challenge},
  year = {2024},
  url1 = {https://github.com/qiskit-community/ibm-quantum-challenge-2024}
}

@misc{qiskit_github,
  author = {{IBM Quantum and Qiskit Developers}},
  title = {Qiskit Open-Source Repositories},
  year = {2023},
  note = {Available at \url{https://github.com/Qiskit}}
}

@misc{qiskit_kaggle,
  author = {Yikai Mao},
  title = {Q-Gen Quantum Circuit Dataset},
  year = {2023},
  url1 = {https://www.kaggle.com/datasets/ykmaoykmao/q-gen-quantum-circuit-dataset}}

@article{qiskit_instruction_data,
  title={QDataSet, quantum datasets for machine learning},
  author={Perrier, Elija and Youssry, Akram and Ferrie, Chris},
  journal={Scientific data},
  volume2={9},
  number2={1},
  pages2={582},
  year={2022},
  publisher2={Nature Publishing Group UK London}
}

@article{Shao_2024,
   title={Survey of Different Large Language Model Architectures: Trends, Benchmarks, and Challenges},
   volume={12},
   ISSN={2169-3536},
   url1={http://dx.doi1.org/10.1109/ACCESS.2024.3482107},
   doi1={10.1109/access.2024.3482107},
   journal={IEEE Access},
   publisher={Institute of Electrical and Electronics Engineers (IEEE)},
   author={Shao, Minghao and Basit, Abdul and Karri, Ramesh and Shafique, Muhammad},
   year={2024},
   pages={188664–188706} }

@article{jiang2024surveylargelanguagemodels,
  title={A survey on large language models for code generation},
  author={Jiang, Juyong and Wang, Fan and Shen, Jiasi and Kim, Sungju and Kim, Sunghun},
  journal={ACM Transactions on Software Engineering and Methodology},
  volume2={35},
  number2={2},
  pages2={1--72},
  year={2026},
  publisher2={ACM New York, NY}
}

@article{zaman2023survey,
  title={A survey on quantum machine learning: Current trends, challenges, opportunities, and the road ahead},
  author={Zaman, Kamila and Marchisio, Alberto and Hanif, Muhammad Abdullah and Shafique, Muhammad},
  journal={arXiv:2310.10315},
  year={2023}
}

@inproceedings{zaman2024po,
  title={PO-QA: A Framework for Portfolio Optimization using Quantum Algorithms},
  author={Zaman, Kamila and Marchisio, Alberto and Kashif, Muhammad and Shafique, Muhammad},
  booktitle={2024 IEEE International Conference on Quantum Computing and Engineering (QCE)},
  volume2={1},
  pages2={1397--1403},
  year={2024},
  organization2={IEEE}
}

@inproceedings{innan2024qfnn,
  title={Qfnn-ffd: Quantum federated neural network for financial fraud detection},
  author={Innan, Nouhaila and Marchisio, Alberto and Bennai, Mohamed and Shafique, Muhammad},
  booktitle={2025 IEEE International Conference on Quantum Software (QSW)},
  pages2={41--47},
  year={2025},
  organization2={IEEE}
}

@inproceedings{el2024quantum,
   title={Quantum Clustering for Cybersecurity},
   url1={http://dx.doi1.org/10.1109/QCE60285.2024.10243},
   doi1={10.1109/qce60285.2024.10243},
   booktitle={2024 IEEE International Conference on Quantum Computing and Engineering (QCE)},
   publisher={IEEE},
   author={Maouaki, Walid El and Innan, Nouhaila and Marchisio, Alberto and others},
   year={2024},
   month=sep, pages={5–10} }

@INPROCEEDINGS{kashif:2022,
  author={Kashif, Muhammad and Al-Kuwari, Saif},
  booktitle={2022 IEEE International Conference on Rebooting Computing (ICRC)}, 
  title={Demonstrating Quantum Advantage in Hybrid Quantum Neural Networks for Model Capacity}, 
  year={2022},
  volume={},
  number={},
  pages2={36-44},
  doi1={10.1109/ICRC57508.2022.00011}}

@article{kashif2023impact,
  title={The impact of cost function globality and locality in hybrid quantum neural networks on nisq devices},
  author={Kashif, Muhammad and Al-Kuwari, Saif},
  journal={Machine Learning: Science and Technology},
  volume={4},
  number={1},
  pages={015004},
  year={2023},
  publisher={IOP Publishing}
}

@INPROCEEDINGS{kashif:qiskit,
  author={Kashif, Muhammad and Al-Kuwari, Saif},
  booktitle={2022 IEEE 19th International Conference on Software Architecture Companion (ICSA-C)}, 
  title={Qiskit As a Simulation Platform for Measurement-based Quantum Computation}, 
  year={2022},
  volume={},
  number={},
  pages2={152-159},
  doi1={10.1109/ICSA-C54293.2022.00037}}

@INPROCEEDINGS{kashif:2025computational,
  author={Kashif, Muhammad and Marchisio, Alberto and Shafique, Muhammad},
  booktitle={2025 62nd ACM/IEEE Design Automation Conference (DAC)}, 
  title={Computational Advantage in Hybrid Quantum Neural Networks: Myth or Reality?}, 
  year={2025},
  volume={},
  number={},
  pages={1-7},
  doi={10.1109/DAC63849.2025.11132906}}

@inproceedings{el2024advqunn,
  title={Advqunn: A methodology for analyzing the adversarial robustness of quanvolutional neural networks},
  author={El Maouaki, Walid and Marchisio, Alberto and Said, Taoufik and Bennai, Mohamed and Shafique, Muhammad},
  booktitle={2024 IEEE International Conference on Quantum Software (QSW)},
  pages={175--181},
  year={2024},
  organization={IEEE}
}

@inproceedings{marchisio2025q,
  title={Q-PORT: Quantum Portfolio Optimization with Resource-Efficient Encoding and Scalability Analysis},
  author={Marchisio, Alberto and Hafeez, Muhammad Umair and Innan, Nouhaila and Kashif, Muhammad and Shafique, Muhammad},
  booktitle={International Conference on Quantum Engineering Sciences and Technologies for Industry and Services},
  pages2={339--347},
  year={2025},
  organization2={Springer}
}

@inproceedings{kashif2025evaluating,
  title={Evaluating quantum amplitude estimation for pricing multi-asset basket options},
  author={Kashif, Muhammad and Khalid, Shaf and Innan, Nouhaila and Marchisio, Alberto and Shafique, Muhammad},
  booktitle={2025 IEEE International Conference on Quantum Artificial Intelligence (QAI)},
  pages={451--458},
  year={2025},
  organization={IEEE}
}

@inproceedings{innan2025quav,
  title={QUAV: Quantum-Assisted Path Planning and Optimization for UAV Navigation with Obstacle Avoidance},
  author={Innan, Nouhaila and Kashif, Muhammad and Marchisio, Alberto and Gan, Yung-Sze and Barbaresco, Frederic and Shafique, Muhammad},
  booktitle={2025 IEEE International Conference on Quantum Artificial Intelligence (QAI)},
  pages2={208--215},
  year={2025},
  organization2={IEEE}
}

@inproceedings{basit2025pennycoder,
  title={PennyCoder: Efficient Domain-Specific LLMs for PennyLane-Based Quantum Code Generation},
  author={Basit, Abdul and Shao, Minghao and Asif, Muhammad Haider and Innan, Nouhaila and Kashif, Muhammad and Marchisio, Alberto and Shafique, Muhammad},
  booktitle={2025 IEEE International Conference on Quantum Computing and Engineering (QCE)},
  volume2={2},
  pages2={229--234},
  year={2025},
  organization2={IEEE}
}

@inproceedings{basit2025qhackbench,
  title={Qhackbench: Benchmarking large language models for quantum code generation using pennylane hackathon challenges},
  author={Basit, Abdul and others},
  booktitle={2025 IEEE International Conference on Quantum Artificial Intelligence (QAI)},
  pages2={316--322},
  year={2025},
  organization2={IEEE}
}
\end{spacing}

\end{document}